\RequirePackage{fix-cm}
\documentclass[twocolumn]{svjour3}
\smartqed
\usepackage[utf8]{inputenc}
\usepackage[T1]{fontenc}
\usepackage[hidelinks]{hyperref}
\usepackage{graphicx}
\usepackage{bm}
\usepackage{mathbbol}
\usepackage{braket}
\usepackage{amssymb,amsmath,amsfonts}
\usepackage{cite}
\usepackage[outercaption]{sidecap}
\newcommand{\up}{\uparrow}
\newcommand{\down}{\downarrow}

\begin{document}

\title{Weak Disorder Enhancing the Production of \\ Entanglement in Quantum Walks}

\author{Alexandre C. Orthey Jr.  \and Edgard P. M. Amorim}

\institute{A. C. Orthey Jr.   \at
Departamento de F\'isica, Universidade do Estado de Santa Catarina, 89219-710, Joinville, SC, Brazil \\
                           \and
           E. P. M. Amorim    \at
Departamento de F\'isica, Universidade do Estado de Santa Catarina, 89219-710, Joinville, SC, Brazil\\
                           \email{edgard.amorim@udesc.br}
}

\date{Received: date / Accepted: date}

\maketitle

\begin{abstract}
We find out a few ways to improve the realization of entanglement between the internal (spin) and external (position) degrees of freedom of a quantum particle, through the insertion of disordered time steps in a one-dimensional discrete time quantum walk in different scenarios. The disorder is introduced by a randomly chosen quantum coin obtained from a uniform distribution among infinite quantum coins or only between Hadamard and Fourier coins for all the time steps (strong disorder). We can also decrease the amount of disorder by alternating disordered and ordered time steps throughout a quantum walk or by establishing a probability $p<0.5$ to pick a Fourier coin instead of a Hadamard one for each time step (weak disorder). Our results show that both scenarios lead to maximal entanglement outperforming the ordered quantum walks. However, these last scenarios are more efficient to create entanglement, because they achieve high entanglement rates in fewer time steps than the former ones. In order to compare distinct disordered cases, we perform an average entanglement by averaging over a large set of initial qubits over time starting from one site (local state) or spread over many neighbor positions following a Gaussian distribution. Some transient behaviors from order to disorder in quantum walks are also evaluated and discussed. Experimental remarks based on available experimental platforms from the literature are made.


\end{abstract}

\section{Introduction} \label{sec:1}

The pioneering work of Aharonov \textit{et al.} \cite{aharonov1993quantum} introduced the quantum version of classical random walks, the quantum random walks or simply, quantum walks (QW) \cite{kempe2003quantum,venegas2012quantum} where the quantum walker is a particle that has a two-level state such as spin $1/2$-like (qubit) sited on one or more discrete positions of a regular one-dimensional lattice. The state of QW is described as a tensor product between spin and position states and it time-evolves through a unitary operator composed of a quantum coin and a conditional displacement operator. The quantum coin operates on the qubit, leaving it in a new superposition of spin states, then the conditional displacement operator shifts the spin up (down) state to the right (left) site. The singular dynamics of QW shows a quadratic gain in the position variance (ballistic spreading) and generates entanglement between the internal (spin) and external (position) degrees of freedom.

QW have been attracting considerable attention in many fields such as physics, computation, engineering and biology \cite{venegas2012quantum}. First studies were focused on algorithmic applications of QW as a quantum search engine \cite{shenvi2003quantum}. However, they also offer some insights to understand the process of human decision-making \cite{busemeyer2006quantum} and the energetic efficiency of photosynthetic systems \cite{engel2007evidence}. They are a possible route to make universal computation \cite{childs2009universal,lovett2010universal}, simulate Dirac-like hamiltonians \cite{chandrashekar2013two} and generate teleportation protocols \cite{wang2017generalized}. QW allow us to contrast quantum-mechanics principles with classical ones, for instance, the violation of Leggett-Garg inequality challenging theories based on classical trajectories \cite{robens2015ideal}. Furthermore, they are robust enough to be implemented in several experimental platforms \cite{wang2013physical}. 

The realization of entanglement in QW is very sensitive to their initial conditions as well as to the quantum coin used in their dynamical evolution. The initial state can be any qubit placed on one position (local state) or spread over many adjacent positions following some kind of distribution function (delocalized state). Previous works showed that the maximal entanglement is achieved when the QW start from two initial qubits in a local state \cite{abal2006quantum,abal2006erratum,salimi2012asymptotic,eryiugit2014time}, and from a continuous set of initial qubits in delocalized states \cite{orthey2017asymptotic}. Other works also demonstrated that delocalized states show a rich variety of behaviors, regarding their dispersion and entanglement in contrast with the local ones \cite{valcarcel2010tailoring,romanelli2010distribution,romanelli2012thermodynamic,romanelli2014entanglement,zhang2016creating,orthey2019connecting,ghizoni2019trojan}. 

Since the long-time entanglement in QW has a strong dependence on the initial coin state, we can obtain the average entanglement by averaging over a large ensemble of initial qubits in order to assess the general features. From this perspective, when some kind of disorder (or noise) introduces different quantum coins in distinct time steps and/or positions, some novel broad features emerge \cite{vieira2013dynamically,vieira2014entangling,zeng2017discrete}. In particular, when the quantum coins change in a truly random way with the same probability to have any coin for all the time steps (strong dynamic disorder), the QW state manifests a diffusive behavior with maximal entanglement between spin and position regardless of the initial state \cite{vieira2013dynamically}. However, for this case, the QW asymptotically reach high rates of entanglement slowly, i.e., after a large number of time steps, taking a time still larger when they start from delocalized states. 

Our main objective here is to find other ways to reach better efficiency to generate maximal entanglement. This article is organized as follows. In Sect. \ref{sec:2} we introduce the mathematical formalism about QW. In Sect. \ref{sec:3} we obtain the average entanglement of ordered and disordered QW over time by averaging over a large set of initial qubits placed on one position (local) or spread over many positions following a Gaussian distribution. We investigate the alternation between disordered and ordered steps in QW and the insertion of steady or time-dependent probabilities to randomly choose one between Hadamard and Fourier coins in the QW. Experimental remarks and some conclusions are outlined in Sect. \ref{sec:4} and Sect. \ref{sec:5} respectively.

\section{Mathematical formalism}\label{sec:2}

\subsection{One-dimensional quantum walk}

The one-dimensional quantum walker is a qubit spread over position states, thus the state of QW belongs to the Hilbert space $\mathcal{H}=\mathcal{H}_C\otimes\mathcal{H}_P$ where $\mathcal{H}_C$ is the qubit (coin) space and $\mathcal{H}_P$ is the position space. The coin space is a complex two-dimensional vector space spanned by two spin states $\{\ket{\up}, \ket{\down}\}$ and the position space is a countably infinite-dimensional vector space spanned by a set of orthonormal vectors $\ket{j}$ and $j\in\mathbb{Z}$ as the discrete positions on a lattice. Then, we can write an initial state as follows,
\begin{equation}
\ket{\Psi(0)}=\sum_{j=-\infty}^{+\infty}\left[a(j,0)\ket{\up}+b(j,0)\ket{\down}\right]\otimes\ket{j},
\label{Psi_0}
\end{equation}
with $\sum_j[|a(j,0)|^2+|b(j,0)|^2]=1$ as the normalization condition and the sum is over all integers.

The dynamical evolution of the QW is unitary in discrete time steps. After $N$ time steps, the state of QW is given by, 
\begin{equation}
\ket{\Psi(N)}=\mathcal{T}\prod_{t=1}^{N}U(t)\ket{\Psi(0)},
\label{norm_cond}
\end{equation}
where $\mathcal{T}$ represents a time-ordered product, and 
\begin{equation}
U(t)=S.[C(t)\otimes\mathbb{1}_P],
\label{U_op}
\end{equation}
is the unitary time evolution operator. From the right to the left, $\mathbb{1}_P$ is the identity operator in the position space $\mathcal{H}_P$, $C(t)$ is a time-dependent quantum coin, and $S$ is the conditional displacement operator.
The quantum coin $C(t)$ operates over the spin states and generates a superposition of them. Since an arbitrary quantum coin $C(t)$ belongs to the $SU(2)$, if an irrelevant global phase is neglected, $C(t)$ can be written as
\begin{equation}
\displaystyle
C(t) = 
\begin{pmatrix}
\sqrt{q(t)} & \sqrt{1-q(t)}e^{i\theta(t)} \\ \sqrt{1-q(t)}e^{i\phi(t)} & -\sqrt{q(t)}e^{i[\theta(t)+\phi(t)]}
\end{pmatrix},
\label{coin_spin}
\end{equation}
with three independent parameters $q(t)$, $\theta(t)$ and $\phi(t)$. The first parameter ranges from $0$ to $1$, and it determines whether the coin is balanced ($q(t)=1/2$) or not ($q(t)\neq1/2$). Both last parameters range from $0$ to $2\pi$, and they control the relative phases between spin states. 

The conditional displacement operator $S$ moves the spin up (down) state to the right (left), then from the site $j$ to the site $j+1$ ($j-1$),
\begin{equation}
S=\sum_j(\ket{\up}\bra{\up}\otimes\ket{j+1}\bra{j}+\ket{\down}\bra{\down}\otimes\ket{j-1}\bra{j}).
\label{S_op}
\end{equation}
This operator generates entanglement between the internal (spin) and external (position) degrees of freedom throughout the dynamical evolution of the QW.

\subsection{Initial conditions}

The initial conditions are given by the initial qubit and position states. First, let us consider an arbitrary initial qubit or coin state,
\begin{equation}
\ket{\Psi_C}=\cos\left(\dfrac{\alpha}{2}\right)\ket{\up}+e^{i\beta}\sin\left(\dfrac{\alpha}{2}\right)\ket{\down},
\label{Psi_s}
\end{equation}
in the Bloch sphere representation \cite{nielsen2010quantum} where $\alpha \in[0,\pi]$ and $\beta \in[0,2\pi]$. We employ two kinds of initial position states, local and delocalized (Gaussian) ones. The local state $\ket{\Psi_L}$ has a delta-like probability distribution and by replacing it in \eqref{Psi_0}, we get a general local initial state given by   
\begin{equation}
\ket{\Psi_L(0)}=\ket{\Psi_C}\otimes\ket{0},
\label{Psi_0_Local}
\end{equation}
and let us consider a Gaussian probability distribution where $\sigma_0$ is the initial dispersion, thus a general initial Gaussian state\footnote{The discrete Gaussian states are defined within $j=-100$ and $100$ centered at $j=0$, the condition of normalization gives an error below $10^{-5}\%$ when $\sigma_0=10$.} is
\begin{equation}
\ket{\Psi_G(0)}=\sum_{j=-\infty}^{+\infty}\ket{\Psi_C}\otimes\dfrac{\text{exp}\left(-j^2/4\sigma_0^2\right)}{(2\pi\sigma_0^2)^{\frac{1}{4}}}\ket{j}.
\label{Psi_0_Gauss}
\end{equation}
Our calculations are performed by averaging over a large set of initial coin states (qubits) starting from these two kinds of initial position states.

\subsection{Average entanglement}

The initial local and Gaussian states are pure and since their time evolution is unitary, $\ket{\Psi_L(t)}$ and $\ket{\Psi_G(t)}$ remain pure for the whole of dynamics. Therefore, the entanglement between spin and position can be evaluated by von Neumann entropy, 
\begin{equation}
S_E(\rho(t))=-\mathrm{Tr}[\rho_C(t)\log_2\rho_C(t)], 
\label{SE}
\end{equation}
of the partially reduced spin (coin) state \cite{bennett1996concentrating}, 
\begin{equation}
\rho_C(t)=\mathrm{Tr}_P[\rho(t)],
\label{rho_C}
\end{equation}
where $\rho(t)=\ket{\Psi(t)}\bra{\Psi(t)}$, $\mathrm{Tr}_P[\cdot]$ is the partial trace over the positions and
\begin{equation}
\rho_C(t)= \begin{pmatrix}
A(t) & \gamma(t) \\
\gamma^*(t) & B(t)
\end{pmatrix},
\end{equation}
with $A(t)=\sum_j|a(j,t)|^2$, $B(t)=\sum_j|b(j,t)|^2$, $\gamma(t)=\sum_ja(j,t)b^*(j,t)$ with $\gamma^*(t)$ being the complex conjugate of $\gamma(t)$. By diagonalizing $\rho_C(t)$, we have
\begin{equation}\label{SE_t}
S_E(\rho(t))=-\lambda_+(t)\log_2\lambda_{+}(t)
-\lambda_{-}(t)\log_2\lambda_{-}(t),
\end{equation}
with eigenvalues,
\begin{equation}
\lambda_{\pm}=\dfrac{1}{2} \pm \sqrt{\dfrac{1}{4}-A(t)(1-A(t))+|\gamma(t)|^2},
\end{equation}
since $A(t)=1-B(t)$. The entropy of entanglement $S_E(t)$ ranges from $0$ for separable states to $1$ for maximal entanglement between spin and position.

The average entanglement for any time step $t$ can be calculated by 
\begin{equation}\label{average_SE}
\braket{S_E(t)}=\sum_{i=1}^n\frac{S_{E,i}(t)}{n},
\end{equation}
where $S_{E,i}(t)$ is the entanglement at a time step $t$, obtained from Eq. \eqref{SE_t} for the QW starting from a $i$-th initial qubit $\ket{\Psi_C}_i$ given by a particular $(\alpha_i, \beta_i)$ over the Bloch sphere. Therefore, after all the $n$ calculations are performed, the average entanglement is obtained.

\subsection{Order, strong disorder and weak disorder}

The ordered QW have a steady quantum coin over all the time steps and positions, such as Hadamard or Fourier (Kempe) coins. Both coins are unbiased ($q=1/2$), and while the Hadamard coin creates a superposition without relative phases between spin states ($\theta=\varphi=0$), the Fourier coin imposes $\pi/2$ ($\theta=\varphi=\pi/2$), then using Eq. \eqref{coin_spin} we have
\begin{equation}
C_{\mathrm{Hadamard}}=\dfrac{1}{\sqrt{2}}\begin{pmatrix}
1 & 1\\
1 & -1
\end{pmatrix}, \qquad
C_{\mathrm{Fourier}}=\dfrac{1}{\sqrt{2}}\begin{pmatrix}
1 & i\\
i & 1
\end{pmatrix}.
\end{equation}

The strongly dynamically disordered (SDD) QW is given by a quantum coin $C(t)$ with $q(t)$, $\theta(t)$ and $\varphi(t)$ for all positions, however $C(t)$ randomly switches for each time step in two distinct ways. The first way, namely SDD$_2$ QW, $C(t)$ is randomly chosen between Hadamard and Fourier coins, thus $q(t)$ is fixed, and $\theta=\varphi=(\pi r_N)/2$ where the random integer $r_N$ is picked $0$ or $1$ with same probability. The second way, namely SDD$_\infty$ QW, $C(t)$ can be any coin from the $SU(2)$, therefore $q(t)=r_q$, $\theta=2\pi r_{\theta}$ and $\varphi=2\pi r_{\varphi}$ such that the random real numbers $r_q$, $r_{\theta}$ and $r_{\varphi}$ are independently chosen from a uniform distribution within $0$ and $1$. Therefore, while the chosen coin for the $SDD_2$ QW is always only one between Hadamard and Fourier, in the $SDD_\infty$ QW there are infinite possibilities to choose a different coin for each time step.

In order to study some situations between ordered QW and SDD QW, we establish weakly dynamically disordered (WDD$_2$) QW by means of a probability $p\in[0,1]$ to have a Fourier coin instead of a Hadamard one at each time step. Notice that, WDD$_2$ QW with $p=0$ and $0.5$ recover the Hadamard QW and SDD$_2$ QW respectively. If $p(t)$ varies from $0$ to $0.5$ we can simulate a transient dynamics from an ordered QW to SDD$_2$ QW and vice versa. 

\section{Entanglement production }\label{sec:3}

\subsection{Order and disorder}

\begin{figure}
\includegraphics[width=0.8\linewidth]{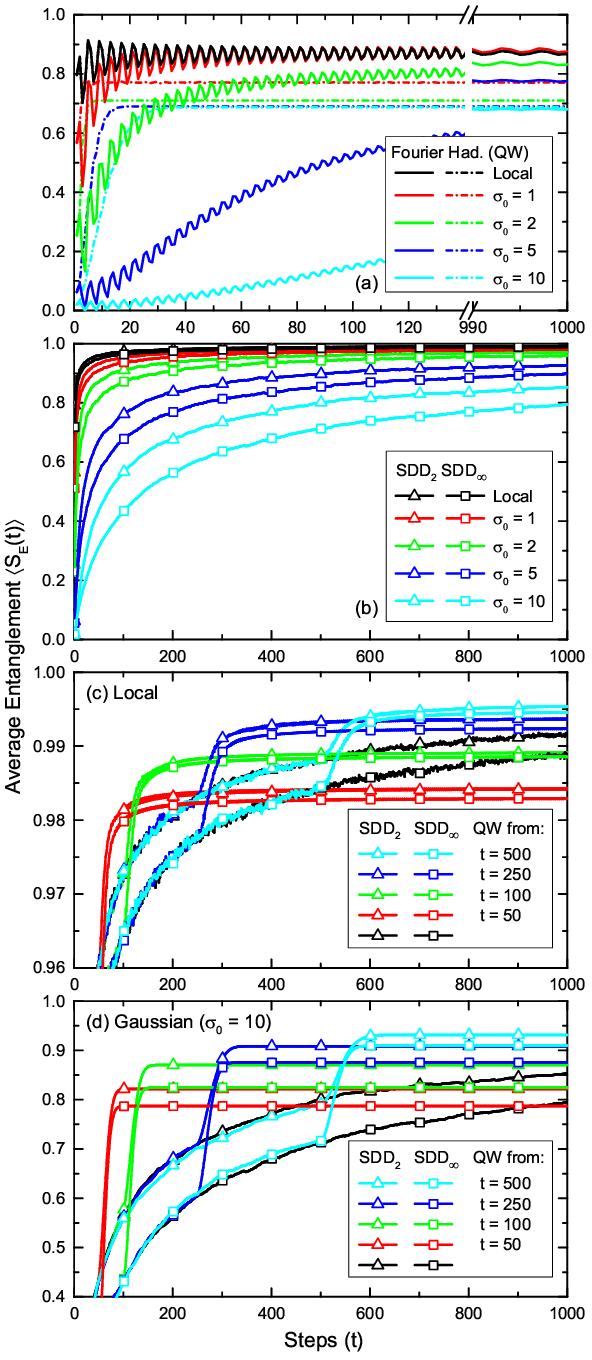}
\centering
\caption{(color online) Average entanglement $\braket{S_E(t)}$ of (a) two ordered QW (Hadamard and Fourier) with a break region between $t=140$ and $990$, (b) $SDD_2$ (triangle) and $SDD_\infty$ QW (square) from local and Gaussian states with $\sigma_0=1$ (red), $2$ (green), $5$ (blue) and $10$ (cyan), (c) $SDD_2$ (triangle) and $SDD_\infty$ QW (square) with an ordered Hadamard QW starting from $t=500$ (cyan), $250$ (blue), $100$ (green) and $50$ (red) from local and (d) Gaussian state with $\sigma_0=10$. Black lines show SDD$_2$ QW and SDD$_\infty$ QW as references.}
\label{fig1} 
\end{figure}

Ordered QW and SDD QW have distinct characteristics regarding their entanglement generation. While the ordered QW reach a maximal entanglement for specific initial states \cite{abal2006quantum,abal2006erratum,salimi2012asymptotic,eryiugit2014time,orthey2017asymptotic}, the SDD QW always achieve the maximal entanglement, but for distinct time rates depending on the initial state \cite{vieira2013dynamically,vieira2014entangling,zeng2017discrete}. Since the entanglement generation of ordered QW and the rate of entanglement production in a disordered context are both sensitive to their initial conditions, in order to properly compare them and observe their general features, we perform the simulations to obtain the average entanglement using Eq. \eqref{average_SE}. The average entanglement is computed for a set of initial qubits from $\alpha=0$ to $\pi$ and $\beta=0$ to $2\pi$ in independent increments of $0.1$, given a total of $2,016$ initial qubits. The numerical simulations are made starting from local and Gaussian position initial states.

Figure \ref{fig1} shows the average entanglement $\braket{S_E}$ over time of (a) two ordered QW (Hadamard and Fourier) and (b) SDD$_2$ QW and SDD$_\infty$ QW starting from local and Gaussian states. Among the ordered cases in Fig. \ref{fig1} (a), while Hadamard and Fourier QW starting from a local state have the same behavior and average long-time entanglement, Hadamard QW starting from delocalized states reach high entanglement rates faster than Fourier ones. All ordered cases exhibit average long-time entanglement $\braket{S_E}<1$ and the more delocalized the initial states are, the lower the values where they achieve the convergence are, with $\braket{S_E}\sim0.688$ as the lower bound for $\sigma_0\gg 1$ \cite{orthey2017asymptotic}. Figure \ref{fig1} (b) indicates that SDD$_2$ QW and SDD$_\infty$ QW lead to $\braket{S_E}\rightarrow 1$ for $t\rightarrow\infty$, then the dynamic disorder is a sufficient condition to generate maximal entanglement \cite{vieira2013dynamically,vieira2014entangling}. It is worth noting that SDD$_2$ QW is more efficient than SDD$_\infty$ QW, since the former takes fewer time steps to reach better rates of average entanglement regardless of initial position states. 

\begin{figure*}
\centering
\includegraphics[width=0.8\linewidth]{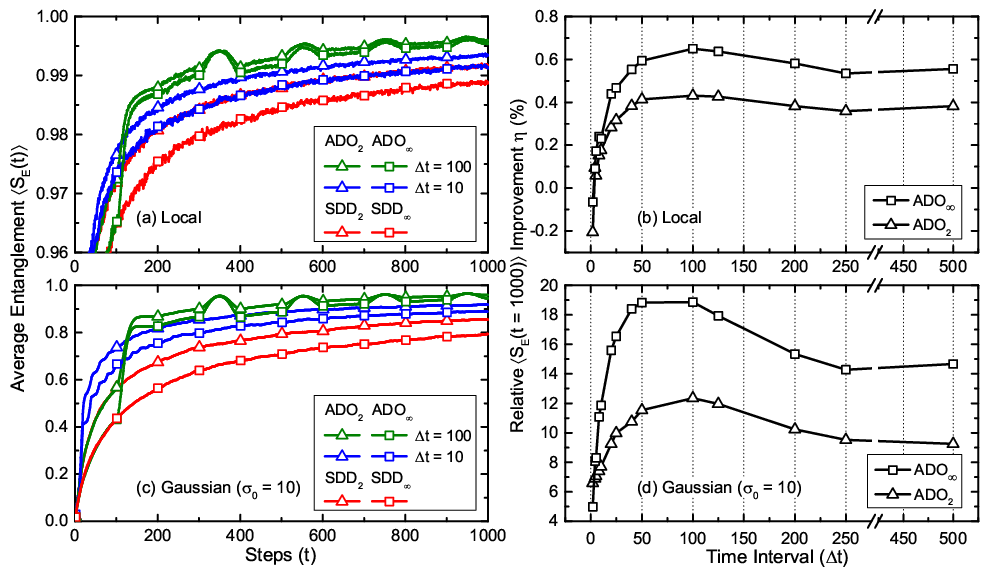}
\caption{(color online) Average entanglement $\braket{S_E(t)}$ of ADO$_2$ in QW (triangle) and ADO$_\infty$ in QW (square) for $\Delta t=100$ (olive) and $\Delta t=10$ (blue) with $SDD_2$ QW (triangle) and $SDD_\infty$ QW (square) both in red starting from (a) local state and (b) Gaussian state with $\sigma_0=10$. Relative average entanglement improvement $\eta (\%)$ calculated for $t=1000$ of ADO$_2$ in QW (triangle) and ADO$_\infty$ in QW (square) with different values of $\Delta t\in[1,500]$ from (c) local state and (d) Gaussian state with $\sigma_0=10$. (c)-(d) The lines are a guide for the eyes.}
\label{fig2}
\end{figure*}

When we start ordered QW from a particular time step of SDD QW, in attempt to recover the ordered behavior, a kink appears in the average entanglement and it reaches a long-time entanglement $\braket{S_E}<1$. Figure \ref{fig1} displays SDD$_2$ QW and SDD$_\infty$ QW from (c) local state and (d) Gaussian state ($\sigma_0=10$) with an ordered QW starting from distinct time steps. The difference between the average entanglement after and before the kink drops asymptotically over time and the time interval to achieve the long-time entanglement increases as the time goes on. This behavior raises the following issue: is there any particular combination of disorder and order, which leads to a maximal entanglement condition in fewer time steps than the SDD$_2$ QW? The next section is devoted to investigate this question.

\subsection{Alternating disorder and order}

The results from the previous section suggest possible alternative routes to achieve high entanglement rates through the insertion of ordered time steps in SDD QW. Let us consider SDD QW with a fixed periodic time interval $\Delta t$ of ordered steps, i.e, if $\Delta t=10$ the first $10$ steps are disordered, followed by the next $10$ ordered steps, and so on. Moreover, we chose all the values $\Delta t$ to assure the same number of ordered and disordered steps in all calculations. In order to make a comparison between the efficiency of alternating disorder and order (ADO) in QW to SDD$_2$ QW and SDD$_\infty$ QW, we define a relative average entanglement improvement $\eta$ as 
\begin{equation}
\eta= \dfrac{\braket{S_E(1000)}_{\mathrm{ADO}}}{\braket{S_E(1000)}_{\mathrm{SDD}}}-1,
\end{equation}
at $t=1000$ when all cases achieve a similar entanglement rate.

Figure \ref{fig2} shows the average entanglement of ADO in QW for $\Delta t=10$ and $100$ and $\eta$ for distinct values of time interval $\Delta t$ starting from (a)-(b) local state and (c)-(d) Gaussian state with $\sigma_0=10$. For all cases depicted in Fig. \ref{fig2} (a) and (c), ADO in QW outperforming the SDD QW for the during the whole evolution of time. Peculiarly, each case of ADO in QW starting from the Gaussian state is notably greater than the local one when we compare them to their respective SDD QW efficiency. On the one hand, ADO$_2$ in QW has a better average entanglement efficiency than ADO$_\infty$ in QW, which is similar to what is evaluated between SDD$_2$ QW and SDD$_\infty$ QW. On the other hand, if we compare the entanglement efficiency of each ADO in QW to their respective SDD QW as shown in Fig. \ref{fig2} (b) and (d), ADO$_\infty$ in QW improves the entanglement attained better than ADO$_2$ in QW regardless of initial state. For both kinds of disorder and initial states, the best improvement of ADO$_2$ (ADO$_\infty$) in QW is achieved with $\Delta t=100$ being $\eta=0.65\%$ ($0.43\%$) starting from local and $18.9\%$ ($12.3\%$) from Gaussian state. The only exception is the case with $\Delta t=2$ (local state) which the average entanglement has a tiny decrease about $0.21\%$ ($0.06\%$) of ADO$_2$ (ADO$_\infty$) in QW respectively.

\subsection{Weak disorder}

SDD QW and ADO in QW with a steady time interval lead to the maximal entanglement condition for $t\rightarrow\infty$, however they have distinct time rates to reach it. Since the introduction of ordered steps among disordered ones improves the entanglement achievement, it seems reasonable to suppose that weak disorder could be more efficient than the strong disorder for creating entanglement. Therefore, a controlled way to decrease the amount of disorder can be reached by introducing a probability to have one coin instead of the other. Once a random choice between Hadamard and Fourier coins has a better efficiency to achieve high entanglement rates than the one among infinite possibilities of coins \cite{vieira2013dynamically,vieira2014entangling} or with two other coins \cite{zeng2017discrete}, then from this point, we will just deal with the first kind of disorder. Let us consider WDD$_2$ QW with a probability $p(t)\in[0,1]$ to obtain a Fourier coin, where for $p=0$ we have a Hadamard QW, for $p=0.5$ we recover SDD$_2$ QW and for $p=1$ we have a Fourier QW.

\begin{figure}
\centering
\includegraphics[width=0.8\linewidth]{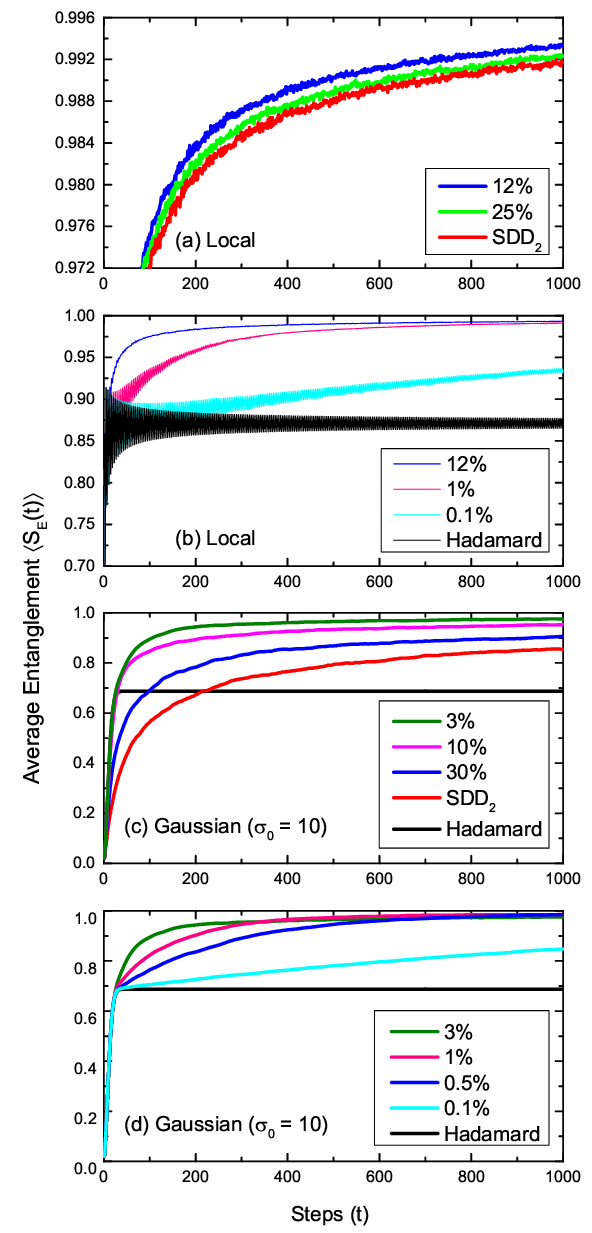}
\caption{(color online) Average entanglement $\braket{S_E(t)}$ of WDD$_2$ QW starting from (a)-(b) local state and (c)-(d) Gaussian state with $\sigma_0=10$ for some values of probability $p$ to obtain a Fourier coin: (a) $12\%$ (blue), $25\%$ (green); (b) $12\%$ (blue), $1\%$ (magenta), $0.1\%$ (cyan); (c) $3\%$ (olive), $10\%$ (pink), $30\%$ (blue); (d) $3\%$ (olive), $1\%$ (magenta), $0.5\%$ (blue), $0.1\%$ (cyan). Red and black lines represent SDD$_2$ QW and Hadamard QW respectively.}
\label{fig3} 
\end{figure}

\begin{figure}
\centering
\includegraphics[width=0.8\linewidth]{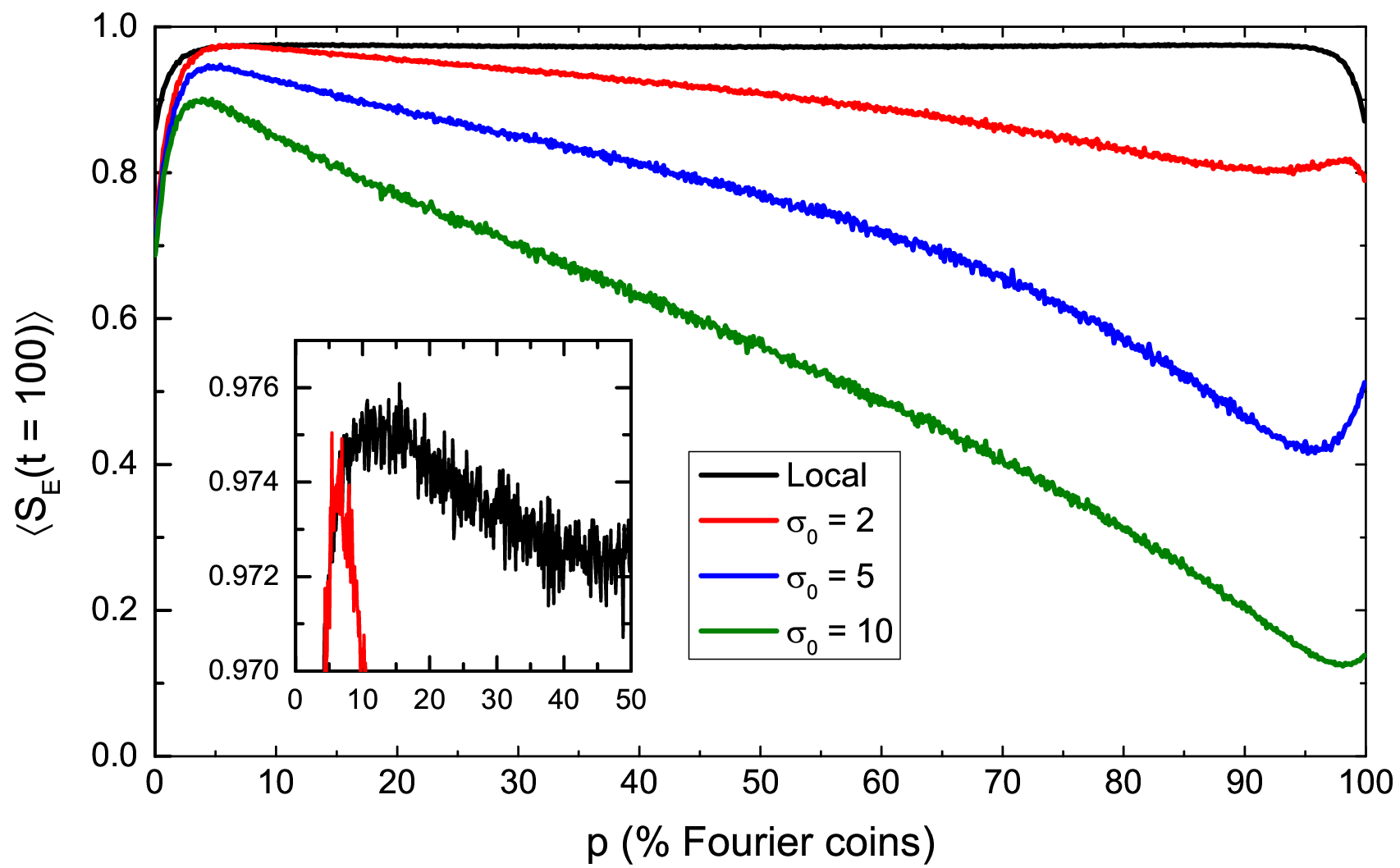}
\caption{(color online) Average entanglement $\braket{S_E(t=100)}$ of WDD$_2$ QW starting from local state (black) and Gaussian states with $\sigma_0=2$ (red), $5$ (blue) and $10$ (olive) for $1000$ different values of $p\in [0,1]$. The inset magnifies upside region.}
\label{fig4} 
\end{figure}

Figure \ref{fig3} shows the average entanglement of WDD$_2$ QW for different values of $p$. WDD$_2$ QW starting from (a)-(b) local state with $p=12\%$ and (c)-(d) Gaussian ($\sigma_0=10$) state with $p=3\%$ are the most effective scenarios to generate entanglement outperforming the entanglement achieved by SDD$_2$ QW across the whole dynamical evolution. Since all cases of WDD$_2$ QW have asymptotic long-time entanglement, in Fig. \ref{fig3} is already possible to distinguish the best entanglement situations from $t=100$. Therefore, in Fig. \ref{fig4} we obtain the $\braket{S_E(t=100)}$ as function of $p$, which reveals the following best $p$ values: $\sim 12\%$ starting from local state and $\sim 9\%$, $\sim 6\%$ and $\sim 3\%$ from Gaussian states with $\sigma_0=2$, $5$ and $10$ respectively.

\begin{figure}
\centering
\includegraphics[width=0.8\linewidth]{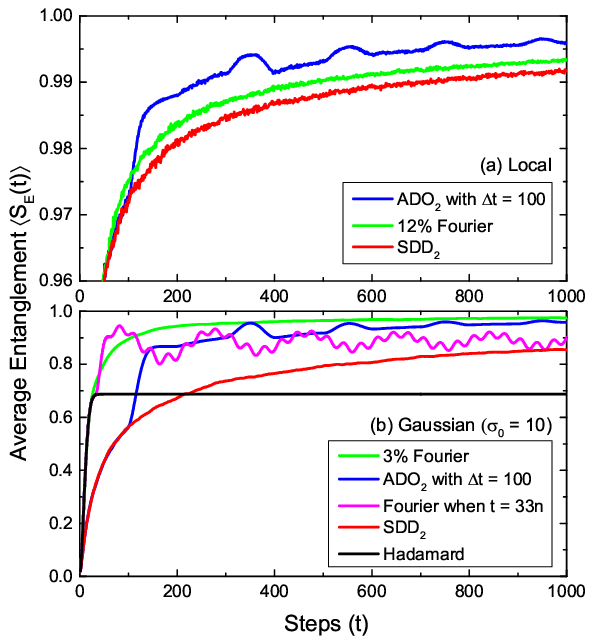}
\caption{(color online) Average entanglement $\braket{S_E(t)}$ starting from (a) local state and (b) Gaussian state with $\sigma_0=10$ of ADO$_2$ in QW with $\Delta t=100$ (blue), WDD$_2$ QW with (a) $p=12\%$ and (b) $3\%$ (green), SDD$_2$ QW (red), (b) Hadamard QW with Fourier coin at $t=33n$ with $n\in\mathbb{N}$ (magenta) and Hadamard QW (black).}
\label{fig5} 
\end{figure}

Figure \ref{fig5} compares the average entanglement behavior for the most notable cases studied here. When the QW start from the local state, ADO$_2$ in QW ($\Delta t=100$) exceeds WDD$_2$ QW with $p=12\%$ and SDD$_2$ QW, while from the Gaussian state, WDD$_2$ QW with $p=3\%$ defeats ADO$_2$ in QW ($\Delta t=100$) and SDD$_2$ QW, in both cases over almost the whole evolution on time. Moreover, we add an ordered simulation with one Fourier coin every $33$ steps, which gives $3\%$ of Fourier coins over all the steps, in order to emphasize the need of disorder to achieve maximal entanglement \cite{vieira2013dynamically,vieira2014entangling}.   

\subsection{Transition between order and disorder}

The transient behavior from order to disorder or from disorder to order can be appraised by means of WDD$_2$ QW with a time-dependent probability $p(t)$. We propose here three smooth analytic functions for $p(t)$ detached in the Tab. \ref{tab1}. They allow us to study a linear and two quadratic (up and down concaves) time-dependent probabilities from Hadamard QW with $p(0)=0$ to SDD$_2$ QW with $p(N)=0.5$ and vice versa over a walk with $N=1000$ time steps.

\begin{figure}
\centering
\includegraphics[width=0.8\linewidth]{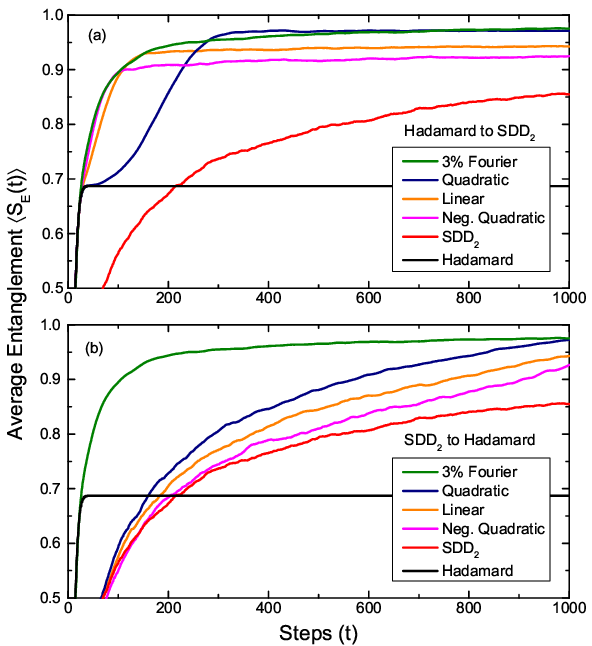}
\caption{Average entanglement $\braket{S_E(t)}$ starting from Gaussian state with $\sigma_0=10$ of the transient behavior (a) from Hadamard QW to SDD$_2$ QW and (b) from SDD$_2$ QW to Hadamard QW by means of WDD$_2$ QW with quadratic (blue), linear (orange) and negative quadratic (magenta) functions for $p(t)$ as showed in the Tab. \ref{tab1}. WDD$_2$ QW with $p=3\%$ (olive), SDD$_2$ QW (red) and Hadamard QW (black) as references.}
\label{fig6} 
\end{figure}

Figure \ref{fig6} shows the average entanglement for all transient scenarios starting from Gaussian state outperforming the SDD$_2$ QW, nonetheless they are worse than WDD$_2$ QW with a steady probability $p=3\%$. Between the three types of time-dependent probabilities, the quadratic overcomes the linear and negative quadratic over most of the time steps. The average entanglement of WDD$_2$ QW with steady probability at $t=1000$ is just $0.4\%$ and $0.3\%$ higher than WDD$_2$ QW with quadratic time-dependent probability for the transitions from Hadamard QW to SDD$_2$ QW and from SDD$_2$ QW to Hadamard QW as depicted in Fig. \ref{fig6} (a) and (b) respectively. To conclude, it is worth mentioning that the evolution from Hadamard QW to SDD$_2$ QW creates higher entanglement in fewer time steps than the opposite case.

\begin{table}[t]
\caption{Expressions for $p(t)$ ($\%$ Fourier coins) from $t=0$ up to $N=1000$: transition from Hadamard QW to SDD$_2$ QW (H$\rightarrow$S) and the opposite (S$\rightarrow$H).}
\label{tab1}
\begin{tabular}{lccc} 
\hline
p(t)              & Linear             & Quadratic               & Negative Quadratic   \\
\hline
H$\rightarrow$S & $ \dfrac{t}{2N}$   & $\dfrac{t^2}{2N^2}$     & $\dfrac{1}{2}-\dfrac{(t-N)^2}{2N^2}$ \\[8pt]
S$\rightarrow$H & $-\dfrac{t-N}{2N}$ & $\dfrac{(t-N)^2}{2N^2}$ & $\dfrac{1}{2}-\dfrac{t^2}{2N^2}$     \\[8pt]
\hline
\end{tabular}
\end{table}

\section{Experimental remarks} \label{sec:4} 

Nowadays, there are available experimental platforms to implement QW with dynamic disorder based on the photon as a quantum walker \cite{wang2013physical}. These platforms are capable to evaluate different disordered scenarios presented here with QW starting from an arbitrary qubit (coin state) with initial local and/or delocalized states. 

Schreiber \textit{et al.} proposed an arrangement with passive optical elements and a fast-switching electro-optical modulator (EOM) able to perform a QW with controllable dynamics \cite{schreiber2010photons,schreiber2011decoherence}. The polarization of photon (vertical and horizontal) and its arrival times at the photodetector are the internal and external degrees of freedom respectively. The initial local state is prepared by using half-wave (HWP) and quarter-wave plates (QWP). The random quantum coin is implemented by another HWP and EOM, while the conditional displacement by two polarizing beam splitters and an extra fiber line responsible for delaying the horizontally polarized light. The temporal difference between both polarization components corresponds to one discrete time step. This technique allows the authors to introduce dynamic disorder with different time scales \cite{schreiber2010photons,schreiber2011decoherence}. Thus, measurements of entanglement of QW with disorder applied in distinct time ranges could experimentally verify our results shown here.

Another platform to implement disordered QW uses integrated waveguide circuits built by femtosecond laser writing techniques with directional couplers playing the role of beam splitters and controlled phase shifts are realized by deforming the waveguides at the output of each directional coupler \cite{crespi2013anderson}. The initial state could be a single photon or two entangled photons whose state can obey the Fermionic or Bosonic statistics. Two perpendicular axes along the circuit indicate the different positions and time steps. The dynamical disorder is introduced by keeping the same phase shifts set for all positions but changing the set from a time step to another \cite{crespi2013anderson}. This experimental platform performs a QW with one photon or more from different input positions. Therefore, in principle, a delocalized initial state could be prepared as an input state. 

The last platform is based on the manipulation of orbital angular momentum (OAM) of photons \cite{zhang2010implementation} from a unique light beam \cite{goyal2013implementing,cardano2015quantum}. The internal degree of freedom is determined by the left or right circular polarization of photons and the external degree of freedom is the $z$-component of OAM of photons. The initial coin state is selected by HWP and QWP, and the position states can be prepared as a local state or a superposition of generic OAM states \cite{cardano2015quantum}. The step is performed by an inhomogeneous and anisotropic birefringent liquid-crystal arranged in a specific pattern (q-plate) \cite{marrucci2006optical}. The dynamical disorder could be implemented by modulating the voltage applied to the q-plates and/or changing the phase-retardation and orientation of the waveplates \cite{cardano2017detection}.

All experimental proposes above have notable advantages in terms of phase stability and scalability, compared to an optical quantum quincunx (Galton's board) \cite{do2005experimental}. However, a measure of entanglement would imply in obtaining the partially reduced coin state $\rho_C(t)$ through a few changes in the experimental schemes above \cite{vieira2013dynamically}. Since a general photon polarization can be written as $\rho_C(t)=\mathbb{1}_C+\sum_{j=1}^{3}r_j\sigma_j$, with $\sigma_j$ being Pauli matrices, it could be experimentally obtained by performing the measure of the average polarization of the photon in the vertical/horizontal axis ($r_3$), in the $\pm 45^o$ axis ($r_1$), and the average right/left circular polarization ($r_2$) \cite{peres2002quantum}. These measurements could be done through an adequate arrangement of HWP and QWP. The collected data is associated to $\rho(t)$, therefore after a post-processing measurement by tracing out its position degrees of freedom, we can get $\rho_C(t)$ \cite{vieira2013dynamically}.

\section{Conclusions} \label{sec:5} 

We studied the interplay between order and dynamic disorder in QW, in attempt to better understand the rules for creating maximal entanglement efficiently. First, in order to cover the general features and perform a fair comparison between all cases investigated, we established an average entanglement calculation given by averaging over a large set of initial coin states over time starting from two variety of position states. Second, since all cases studied led to the maximal entanglement, we searched for those scenarios which achieved high entanglement rates more rapidly. Third, we reviewed the average entanglement behavior of ordered QW (Hadamard) and SDD QW based on recent literature \cite{vieira2013dynamically,vieira2014entangling} as the starting point for our study.

We carried out some calculations and realized that the change from SDD QW to Hadamard QW introduced an abrupt increase (kink) in the entanglement, but the long-time entanglement converged to values below the maximal value. From this point, our main aim was to find out other routes combining order and disorder to achieve high entanglement using fewer time steps than in SDD QW. We simulated a periodic alternation between SDD QW and Hadamard QW for different time intervals and we introduced WDD QW with a steady probability to pick one coin between Hadamard and Fourier for each time step. Moreover, we studied the transient behavior from Hadamard QW to SDD QW and the opposite situation by means of a time-dependent probability along the walk.

Our main findings regarding the efficient creation of entanglement could be summarized in the following statements: (i) a truly random choice between Hadamard and Fourier coins over all the time steps leads to a maximal entanglement more readily than the one among infinite coins, which indicates a dependence on the variation of the used coins over time, once both $\theta$ and $\varphi$ have $\pi/2$ of phase difference between Hadamard and Fourier coins \cite{vieira2013dynamically,vieira2014entangling,zeng2017discrete}; (ii) for most of the cases, ADO in QW with a constant time interval $\Delta t$ exhibited a better efficiency than SDD QW; (iii) the efficiency to generate entanglement depends not only on the number of ordered or disordered steps, but also on how they are employed in the QW; (iv) the transient behavior from Hadamard QW to SDD$_2$ QW has a better efficiency to generate entanglement than the contrary case. 

We also found out some similarities and differences starting from each position initial state: (i) the convergence to the long-time entanglement of QW starting from a Gaussian state is slower than from a local one \cite{orthey2017asymptotic} for all disordered cases; (ii) ADO in QW with $\Delta t=100$ is the best case from both position states, being the relative efficiency differences between ADO in QW and SDD QW more significant starting from a Gaussian state; (iii) regarding WDD$_2$ QW to create entanglement, the probability $p$ increases with the localization of the state, starting in $3\%$ from Gaussian state with $\sigma_0=10$ up to $12\%$ from local state; (iv) ADO in QW with $\Delta t=100$ and WDD$_2$ QW with a constant probability $p=3\%$ are the most efficient cases for creating entanglement starting from local and Gaussian states respectively. 

Finally, since the quantum entanglement is a key issue in the quantum information processing with implications in developing quantum algorithm protocols, quantum teleportation, quantum cryptography, dense coding and for building a quantum computer \cite{horodecki2009quantum}, we believe the efficient creation of entanglement could be an important issue in this context. Moreover, our achievements raise novel further questions. For example, is it possible to observe this enhancement of entanglement through dynamical disorder for two- or three-dimensional QW \cite{chandrashekar2013disorder}? And also, for two or more quantum walkers? What should be the interplay between dynamical and static disorder to generate high entanglement with localization \cite{vieira2013dynamically,vieira2014entangling,chandrashekar2013disorder}? We hope our findings inspire other investigations on this subject and the experimental researchers can test our results for diverse platforms.

\begin{acknowledgements}
This study was financed in part by the Coordena\c{c}\~ao de Aperfei\c{c}oamento de Pessoal de N\'ivel Superior - Brasil (CAPES) - Finance Code 001. We thank F. Cardano for kindly enlightening some experimental aspects mentioned here and also J. Longo for her careful reading and suggestions to improve the presentation of the manuscript.
\end{acknowledgements}

\end{document}